\documentclass[aps,prl,twocolumn,superscriptaddress]{revtex4}
\usepackage{color}
\usepackage{amssymb}
\usepackage{epsfig}
\usepackage{graphicx,amsmath}
\begin{document}

\title{Comment on "Topological excitations and the
dynamic structure factor of spin liquids on the kagome lattice"
(Punk, M., Chowdhury, D. \& Sachdev, S. Nature Physics {\bf 10},
289-293 (2014))}

\author{V. R. Shaginyan}\email{vrshag@thd.pnpi.spb.ru} \affiliation{Petersburg
Nuclear Physics Institute,  Gatchina, 188300,
Russia}\affiliation{Clark Atlanta University, Atlanta, GA 30314,
USA} \author{M. Ya. Amusia}\affiliation{Racah Institute of Physics,
Hebrew University, Jerusalem 91904, Israel}\affiliation{Ioffe
Physical Technical Institute, RAS, St. Petersburg 194021, Russia}
\author{J.~W.~Clark}
\affiliation{McDonnell Center for the Space Sciences \& Department
of Physics, Washington University, St.~Louis, MO 63130,
USA}\author{G.~S.~Japaridze}\affiliation{Clark Atlanta University,
Atlanta, GA 30314, USA}
\author{A. Z. Msezane}\affiliation{Clark Atlanta University, Atlanta, GA 30314,
USA}\author{K. G. Popov}\affiliation{Komi Science Center, Ural
Division, RAS, Syktyvkar, 167982, Russia}\author{V. A.
Stephanovich} \affiliation{Opole University, Institute of Physics,
Opole, 45-052, Poland}\author{M. V. Zverev} \affiliation{Russian
Research Center Kurchatov Institute, Moscow, 123182,
Russia}\affiliation{ Moscow Institute of Physics and Technology,
Moscow, 123098, Russia}
\author{V. A. Khodel}
\affiliation{Russian Research Center Kurchatov Institute, Moscow,
123182, Russia} \affiliation{McDonnell Center for the Space
Sciences \& Department of Physics, Washington University,
St.~Louis, MO 63130, USA} 
\maketitle

{\bf Letter to the Editor} --- The authors of a recent paper
\cite{nat14} evidently take the view that the whole of progress
made toward a theoretical understanding of the physics of
quantum spin liquids (QSL) is associated with models of the kind
proposed and applied in their present work.  As motivation for
this work, they observe that in contrast to existing theoretical
models of both gapped and gapless spin liquids, which give rise
to sharp dispersive features in the dynamic structure factor,
the measured dynamic structure factor reveals an excitation
continuum that is remarkably flat as a function of frequency.
They go on to assert that ``so far, the only theoretical model
for a spin liquid state on the kagome lattice which naturally
gives rise to a flat excitation band at low energies consists of
the $Z_2$ spin liquids", cited as references 2 to 4.

Here we point out that there already exists a different and
demonstrably successful approach to the QSL problem that does
naturally feature a flat band. This alternative approach is
based on the theory of fermion condensation \cite{fc}, which is
concerned with special solutions admitted by the equations of
the original Landau theory. Such solutions exhibit a fermion
condensate (FC), composed of the totality of single-particle
states belonging to a portion of the spectrum that is completely
flat in some region of momentum space \cite{jetpl}. It has been
shown by Volovik \cite{vol} that solutions having a FC belong to
a different topological class of Fermi liquids than the
conventional ones, even without the visons introduced by the
authors of \cite{nat14} as "topology carriers". Importantly, the
effective mass of a Landau quasiparticle now acquires a
dependence on external parameters. This further development of
the original Landau theory provides for studies of
characteristic non-Fermi-liquid behavior of FC-hosting systems
that are associated with dramatic change of the single-particle
spectrum under tiny variations of input parameters, including
temperature, magnetic field, pressure, doping, etc. In dealing
with QSLs, such an approach allows one to gain insight into the
physical mechanism operative in herbertsmithite, within the much
broader context of similar phenomena in heavy-fermion (HF)
metals and quasicrystals. Notably, it permits one to calculate
the thermodynamic and relaxation properties of QSL as functions
of temperature $T$ and magnetic field $B$. The crux of the
matter within the Landau-theoretic framework is that the kagome
lattice of herbertsmithite supports a strongly correlated QSL
located at the fermion-condensation quantum phase transition
point \cite{fc,prb,my}.  The consequences of this conclusion are
in good agreement with recent experimental data.  Moreover, the
unusual behavior exhibited in QSL is seen to have universal
character, shared with that observed in such different strongly
correlated fermionic systems as HF metals (including the quantum
critical metal $\rm Sr_3Ru_2O_7$, 2D $^3$He) and quasicrystals
\cite{fc,srruo,qc}.

The FC theory gives a consistent, even quantitative, account of
many of the physical properties of quantum spin liquids. In
particular, it allows for calculation of the temperature- and
magnetic-field dependence of their specific heat, dynamic
magnetic susceptibility $\chi$ and reveals their scaling
behavior \cite{prb,my}. The theory relates such different
phenomena as the QSL heat conductance in a magnetic field with
the magnetoresistance in HF metals, as well as with the spin
relaxation rate in QSL and HF metals \cite{epl,pla}. Moreover,
FC theory can describe the imaginary part of $\chi$, exposing
its scaling properties, which have been shown to be similar to
those found in HF metals. The imaginary part of the dynamic
susceptibility is of special importance since it can be directly
measured in neutron scattering experiments \cite{pla}. Thus, the
spin excitations in $\rm ZnCu_3(OH)_6Cl_2$ exhibit the same
behavior as electron excitations of the HF metal and, therefore
form a continuum.
Based on the good agreement with experiment
achieved in the cited theoretical studies, the FC theory
predicts the results of the key experiment of Han et al.\
\cite{han} showing that the excitations in the spin-liquid state
of the kagome antiferromagnet form a dispersionless continuum
\cite{pla}.

We were unable to check the results of the paper \cite{nat14}
because the presentation is fragmentary and lacks calculated
results for the observable physical properties of QSL addressed
above within FC theory.  By contrast, within the context of FC
theory, the presence of a flat band in herbertsmithite was
predicted \cite{prb,my} before its actual experimental discovery
\cite{han}. All these results stand in vivid contrast with those
of the paper \cite{nat14}, where only a qualitative analysis of
the dynamic structure factor $S({\bf k},\omega)$ is presented.
Finally, as should already be apparent, references related to FC
theory are absent from \cite{nat14}.

\end{document}